\documentclass[a4paper,11pt]{article}%
\usepackage{amsfonts}
\usepackage{graphicx}
\usepackage{amsmath}
\usepackage{dsfont}
\usepackage{amssymb}
\usepackage{bm, lmodern}
\usepackage{latexsym}
\usepackage{subcaption}
\usepackage[colorlinks]{hyperref}
\usepackage{color}
\usepackage{enumerate}
\usepackage{enumitem}
\usepackage{cite}
\usepackage{makeidx}
\usepackage{colortbl}%
\newcommand{\bra}{\begin{array}}
\newcommand{\era}{\end{array}}
\newcommand{\beq}{\begin{equation}}
\newcommand{\eeq}{\end{equation}}
\newcommand{\bqr}{\begin{eqnarray}}
\newcommand{\eqr}{\end{eqnarray}}

\def\BC{\bb C}
\def\_\BC{\bbi C}

\def\( {\left(}
\def\) {\right)}
\def\r4{\color{red}}
\def\no2 {{\textstyle{n\over 2}}}

\newcommand{\lb}{\label}

\begin{document}
\begin{titlepage}
\setcounter{page}{1}
\renewcommand{\thefootnote}{\fnsymbol{footnote}}
\begin{flushright}
\end{flushright}
\vspace{5mm}
\begin{center}
{\Large \bf {Entanglement in Three Coupled Harmonic Oscillators}}

\vspace{5mm}

{\bf Abdeldjalil Merdaci\footnote{\sf amerdaci@kfu.edu.sa}}$^{a}$
and
{\bf Ahmed Jellal\footnote{\sf 
a.jellal@ucd.ac.ma}}$^{b}$

\vspace{5mm}

{$^a$\em  Department of Physics, College of Science, King Faisal University,\\
PO Box
380, Alahsa 31982, Saudi Arabia}

{$^{b}$\em Laboratory of Theoretical Physics,  
Faculty of Sciences, Choua\"ib Doukkali University},\\
{\em PO Box 20, 24000 El Jadida, Morocco}

\vspace{30mm}

\begin{abstract}
We develop an approach in solving exactly the problem of three-body
oscillators including general quadratic interactions in the coordinates
for arbitrary masses and couplings. We  introduce a unitary
transformation of three independent angles to end up with
a diagonalized Hamiltonian.
Using the representation theory of the group $SU(3)$, we 
explicitly determine  the solutions of the energy
spectrum. Considering the ground state together with reduced density matrix, we
derive  the corresponding purity function 
that is giving rise to minimal and maximal entanglement under suitable conditions.
The cases of realizing one variable among three is discussed and
know results in literature are recovered.
\end{abstract}

\end{center}
\vspace{3cm}



\end{titlepage}


\section{Introduction}


The entanglement has no classical analog as spin for quantum particles,
which was initiated for the first time by Schr\"odinger  when he introduced
his thought experiment, known as Schr\"odinger cat, to describe the flawed
interpretation of quantum superposition  \cite{1}. 
Based on the assumption of local
realism
Einstein-Podolsky-Rosen (EPR) proposed a theory, called EPR paradox, 
involving
 two spatially separated particles,
which have both perfectly correlated positions and momenta \cite{2}. Bell formulated
 the idea of EPR 
mathematically and showed to be incompatible with the statistical predictions of quantum mechanics  by
proposing a more stringent
tests dealing with a different set of measurements \cite{3}.
In the last decades, entanglement has gained
a
renewed interest mainly because of the development of the quantum information
science \cite{4}. 
In addition, several quantum protocols
like teleportation and quantum dense coding as well as others 
\cite{Ben1,Ben2,Eckert,Murao,Fuchs,Gottesman} are exclusively
realized using
the entangled states.

Entanglements of many  particles (three or more) 
are fascinating quantum systems,
especially when the entanglement is maximal \cite{Zeilinger}. 
In fact, from a large entangled state of many parts of the system
and performing some measurements on certain parts of such state it turns out that
one can get some information about
the state of the rest of the system. 
In general, it is not easy to analyze the entanglement of three or more particles
because of the complicity of the problem. Indeed,
for a system of two qubits,
it is easy to decide whether the system is entangled or not 
and here the
positivity of the partial trace is a necessary and sufficient condition for
separability. However
for a system of three qubits described by the states $(\psi_a, \psi_b, \psi_c)$ things start to be little
bit complicated, because one has to consider 
three bipartitions of the whole system  
and look at their separabilities.  
Now, it may happen that the
state $\psi_a$ will be separable with respect to the first  
partition but not to the last partition of the 
state $\psi_c$.
One might even find that all three partial traces are positive but
the separability  of all three quibits.
In principle, one can have states completely inseparable,
separable with respect to one or two bipartitions, states separable with
respect to all three bipartitions but not completely separable, and fully
separable states. {In addition,
different classification schemes have been developed on the separability
properties of three-mode Gaussian states  in particular that has been done in \cite{Giedke} by deriving  a necessary and sufficient condition to classify Gaussian states.
}

We are interested to the entanglement of three-body system and before doing so, let us mention some relevant works. 
Indeed, a general scheme and realizable procedures for generating three particle
entanglements out of just two pairs of entangled particles from independent
emissions was studied \cite{Zeilinger}. 
The dynamics of mixedness and entanglement was examined by solving the
time-dependent Schr\"odinger equation for three coupled harmonic
oscillator system with arbitrary time-dependent frequency and coupling
constants parameters, assuming that part of oscillators is inaccessible and
remaining oscillators accessible \cite{Park}. 
{An analysis of three
coupled oscillators composed of three-mode interaction (Stokes, anti-Stokes
and phonon) was developed using the representation theory of a Lie algebra
\cite{Abdalla}. 
Experimentally, quantum-mechanical entanglement of three  particles
has been realized \cite{Bouwmeester, Rauschenbeutel, Ferraro} and for more than
three modes we refer to \cite{Pan, Sackett, Bondani, Zhao}.
}

Motivated by \cite{Zeilinger,Park, Abdalla}, we
consider a system of three coupled harmonic oscillators and study the entanglement.  
After
some transformations and using the $SU(3)$ representation, we solve the
eigenvalue equation to obtain the solutions of the energy spectrum.
By extracting the ground state wavefunction, we determine the corresponding
reduced density matrix and then the purity function in terms of the physical parameters. 
 We show that our system can be minimally and maximally entangled 
 under suitable conditions.
To show the validity of
our findings, we recover already published significant results concerning two coupled harmonic oscillators
\cite{Jellal11}. This will be done by distinguishing three different cases according to 
choices of the frequency limits and  adequate variables.

The present paper is organized as follows. In section 2, we consider the
Hamiltonian of three coupled harmonic oscillators
and introduce relevant
transformations. To get the eigenvalues and eigenstates, we use the
representation theory of the group $SU(3)$ in section 3. Section 4 deals with
the entanglement in ground state, which is done by calculating the purity
function and discuss its minimal and maximal values. In section 5, we study  interesting limits {of our results} to show the relevance of
our results. We conclude our results in the final section and finish with two {\bf Appendices}
({\bf A}: for representation of $SU(3)$, {\bf B}: for the two remaining limits).


\section{Hamiltonian model} 


We consider a system  of three coupled
harmonic oscillators of different masses $(m_{1},m_{2},m_{3})$, frequencies
$(\omega_{1},\omega_{2},\omega_{3})$ and couplings $(D_{12},D_{13},D_{23})$.  
It is
described by the  Hamiltonian
\begin{equation}\lb{ham1}
H_{1}=\frac{1}{2}\left(  \frac{p_{1}^{2}}{m_{1}}+\frac{p_{2}^{2}}{m_{2}}%
+\frac{p_{3}^{2}}{m_{3}}+m_{1}\omega_{1}^{2}x_{1}^{2}+m_{2}\omega_{2}^{2}%
x_{2}^{2}+m_{3}\omega_{3}^{2}x_{3}^{2}+D_{12}x_{1}x_{2}+D_{13}x_{1}%
x_{3}+D_{23}x_{2}x_{3}\right)
\end{equation}
which can be written as 
\begin{eqnarray}
H_{2}  & =&\frac{1}{2m}\left(  {P_{1}^{2}}+{P_{2}^{2}}+{{P}_{3}^{2}}\right)
+\frac{m}{2}\left(  \omega_{1}^{2}X_{1}^{2}+\omega_{2}^{2}X_{2}^{2}+\omega
_{3}^{2}X_{3}^{2}\right)  \nonumber\label{ham2}\\
&& +m\left(  J_{12}X_{1}X_{2}+J_{13}X_{1}{X}_{3}+J_{23}X_{2}{X}_{3}\right)
\end{eqnarray}
after
rescaling the phase space variables
\beq
\left(
\begin{matrix}
x_{1}\\
x_{2}\\
x_{3}%
\end{matrix}
\right)= \left(
\begin{matrix}
\mu_{1}^{-1} X_{1}\\
\mu_{2}^{-1} X_{2}\\
\mu_{3}^{-1} X_{3}%
\end{matrix}
\right), \qquad
\left(
\begin{matrix}
p_{1}\\
p_{2}\\
p_{3}%
\end{matrix}
\right)= \left(
\begin{matrix}
\mu_{1} P_{1}\\
\mu_{2} P_{2}\\
\mu_{3} P_{3}%
\end{matrix}
\right)
\eeq
where the involved parameters are given by
\beq
m=\left(  m_{1}m_{2}m_{3}\right)  ^{\frac{1}{3}},\qquad \mu
_{i}=\left(  \frac{m_{i}}{m}\right)  ^{\frac{1}{2}},\qquad \mu_{1}\mu_{2}\mu
_{3}=1,\qquad J_{ij}=\frac{D_{ij}}{2\sqrt{m_{i}m_{j}}}
\eeq
and the indices $i,j=1,2,3$.

Since $H_{2}$ involves
interacting terms, then a straightforward investigation of the basic features
of the system is not an easy task. Nevertheless, one can overcome such situation by
writing the Hamiltonian in matrix form
\begin{equation}
H_{2}=\frac{1}{2m}\sum_{i,j=1}^{3}P_{i}\delta_{ij}P_{j}+\frac{1}{2}%
m\sum_{i,j=1}^{3}X_{i}\mathcal{R}_{ij}X_{j}%
\end{equation}
where the two matrices
take the forms
\begin{equation}
  \mathcal{R}_{ij}  =\left(
\begin{matrix}
\omega_{1}^{2} & J_{12} & J_{13}\\
J_{12} & \omega_{2}^{2} & J_{23}\\
J_{13} & J_{23} & \omega_{3}^{2}%
\end{matrix}
\right)  ,\qquad X\mathbf{=}\left(
\begin{matrix}
X_{1}\\
X_{2}\\
X_{3}%
\end{matrix}
\right).
\end{equation}
Next we proceed by making more transformations to get a Hamiltonian for three
decoupled harmonic oscillators and therefore explicitly determine the
solutions of the energy spectrum. This will be done by making use of an algebraic approach based on the group $SU(3)$.


\section{Group $SU(3)$ and diagonalization}


We will now see how to use a faithful matrix representation to diagonalize the
Hamiltonian $H_{2}$ by introducing the generators of Lie group $SU(3)$, see
{\bf Appendix A}. To start 
let us write the
matrix $\mathcal{R}_{ij}$ in terms of the generators $\lambda_{i}$ 
\begin{equation}
\label{Rij}
\mathcal{R}_{ij}  =J_{12}\lambda_{1}+J_{13}%
\lambda_{4}+J_{23}\lambda_{6}+\text{diag}\left(  \omega_{1}^{2},\omega_{2}%
^{2},\omega_{3}^{2}\right)
\end{equation}
and make 
a rotation with three angles $(\varphi,\phi,\theta)$%
\begin{equation}
\left(
\begin{matrix}
X_{1}\\
X_{2}\\
X_{3}%
\end{matrix}
\right)  = M 
\left(
\begin{matrix}
q_{1}\\
q_{2}\\
q_{3}%
\end{matrix}
\right), \qquad
\left(
\begin{matrix}
P_{1}\\
P_{2}\\
p_{3}%
\end{matrix}
\right)  = M 
\left(
\begin{matrix}
\tilde P_{1}\\
\tilde P_{2}\\
\tilde P_{3}%
\end{matrix}
\right)
\label{244}%
\end{equation}
such that the matrix is of the form
\beq
M=e^{i\varphi\lambda_{7}}e^{i\phi\lambda_{2}}e^{i\theta
\lambda_{5}}
\eeq
and explicitly reads as
\begin{equation}
M=\allowbreak\left(
\begin{array}
[c]{ccc}%
\cos\theta\cos\phi & \sin\phi & \cos\phi\sin\theta\\
-\sin\theta\sin\varphi-\cos\theta\cos\varphi\sin\phi & \cos\phi\cos\varphi &
\cos\theta\sin\varphi-\sin\theta\cos\varphi\sin\phi\\
-\sin\theta\cos\varphi+\cos\theta\sin\phi\sin\varphi & -\cos\phi\sin\varphi &
\cos\theta\cos\varphi+\sin\theta\sin\phi\sin\varphi
\end{array}
\right).\lb{MM}
\end{equation}
Using 
\eqref{Rij} together with \eqref{MM},  
we show the relation
\begin{equation}
 \mathcal{R}_{ij}  =M\text{{diag}}\left(  \Sigma_{1}^{2}%
,\Sigma_{2}^{2},\Sigma_{3}^{2}\right)  M^{-1}
\end{equation}
with the quantities
\begin{align}
\omega_{1}^{2}  &  =\left(  \Sigma_{1}^{2}\cos^{2}\theta+\Sigma_{3}^{2}%
\sin^{2}\theta\right)  \cos^{2}\phi+\Sigma_{2}^{2}\sin^{2}\phi\label{o11}\\
\omega_{2}^{2}  &  =2\tfrac{\left(  \Sigma_{2}^{2}\cos^{2}\phi+\left(
\Sigma_{1}^{2}\cos^{2}\theta+\Sigma_{3}^{2}\sin^{2}\theta\right)  \sin^{2}%
\phi\right)  \cos^{2}\varphi+\left(  \Sigma_{3}^{2}\cos^{2}\theta+\Sigma
_{1}^{2}\sin^{2}\theta\right)  \sin^{2}\varphi}{2}+\tfrac{\Sigma_{1}%
^{2}-\Sigma_{3}^{2}}{2}\sin2\theta\sin\phi\sin2\varphi\label{o22}\\
\omega_{3}^{2}  &  =2\tfrac{\left(  \Sigma_{3}^{2}\cos^{2}\theta+\Sigma
_{1}^{2}\sin^{2}\theta\right)  \cos^{2}\varphi+\left(  \Sigma_{2}^{2}\cos
^{2}\phi+\left(  \Sigma_{1}^{2}\cos^{2}\theta+\Sigma_{3}^{2}\sin^{2}%
\theta\right)  \sin^{2}\phi\right)  \sin^{2}\varphi}{2}-\tfrac{\Sigma_{1}%
^{2}-\Sigma_{3}^{2}}{2}\sin2\theta\sin\phi\sin2\varphi\label{o33}\\
J_{12}  &  =-\tfrac{\left(  \Sigma_{1}^{2}-\Sigma_{2}^{2}\right)  \cos
^{2}\theta+\left(  \Sigma_{3}^{2}-\Sigma_{2}^{2}\right)  \sin^{2}\theta}%
{2}\sin2\phi\cos\varphi-\frac{\Sigma_{1}^{2}-\Sigma_{3}^{2}}{2}\sin2\theta
\cos\phi\sin\varphi\label{j12}\\
J_{13}  &  =\tfrac{\left(  \Sigma_{1}^{2}-\Sigma_{2}^{2}\right)  \cos
^{2}\theta+\left(  \Sigma_{3}^{2}-\Sigma_{2}^{2}\right)  \sin^{2}\theta}%
{2}\sin2\phi\sin\varphi-\frac{\Sigma_{1}^{2}-\Sigma_{3}^{2}}{2}\sin2\theta
\cos\phi\cos\varphi\label{j13}\\
J_{23}  &  =\tfrac{\Sigma_{1}^{2}-\Sigma_{3}^{2}}{2}\sin2\theta\sin\phi
\cos2\varphi-\tfrac{\Sigma_{2}^{2}\cos^{2}\phi+\left(  \Sigma_{1}^{2}\cos
^{2}\theta+\Sigma_{3}^{2}\sin^{2}\theta\right)  \sin^{2}\phi-\Sigma_{3}%
^{2}\cos^{2}\theta-\Sigma_{1}^{2}\sin^{2}\theta}{2}\sin2\varphi\label{j23}.
\end{align}
From these, we map 
the Hamiltonian $H_{2}$ as follows 
\begin{equation}
H_{3}=\frac{1}{2m}\left( \tilde P{_{1}^{2}}+ \tilde
P{_{2}^{2}}+ \tilde P{_{3}^{2}}\right)
+\frac{m}{2}\left(  \Sigma_{1}^{2}q_{1}^{2}+\Sigma_{2}^{2}q_{2}^{2}+\Sigma
_{3}^{2}q_{3}^{2}\right)  \label{ham4}%
\end{equation}
meaning that our system becomes decoupled
(three decoupled harmonic oscillators) 
and
therefore the energy spectrum can be easily obtained. One more thing, the
result obtained by studying quantum propagator for some classes of
three-dimensional three-body systems \cite{dutra} can be recovered from our
results just by taking the special case $\Sigma_{3}\longrightarrow0$.
In fact,
one of the three oscillators becomes a free particle and
then the Hamiltonian can be diagonalized using tow rather than three free parameters.

By introducing the new set of parameters $\rho$, $\varsigma$ and $\kappa$
\begin{equation}\lb{177}
\varpi=\left(  \Sigma_{1}\Sigma_{2}\Sigma_{3}\right)  ^{\frac{1}{3}},\qquad
e^{\varsigma-\rho}=\frac{\Sigma_{1}}{\varpi},\qquad e^{\kappa-\varsigma}%
=\frac{\Sigma_{2}}{\varpi},\qquad e^{\rho-\kappa}=\frac{\Sigma_{3}}{\varpi}%
\end{equation}
we write the Hamiltonian \eqref{ham4} as
\begin{equation}
H_{3}=\frac{1}{2m}\left(  \tilde P{_{1}^{2}}+\tilde P{_{2}^{2}}+\tilde P{_{3}^{2}}\right)
+\frac{m}{2}\varpi^{2}\left(  e^{2\left(  \varsigma-\rho\right)  }q_{1}%
^{2}+e^{2\left(  \kappa-\varsigma\right)  }q_{2}^{2}+e^{2\left(  \rho
-\kappa\right)  }q_{3}^{2}\right)
\end{equation}
which is now describing three decoupled harmonic oscillators with the same mass $m$ and three different frequencies. Then solving 
the eigenvalue
equation
\begin{equation}
H_{4}\mid n_{1},n_{2},n_{3}\rangle=E_{n_{1},n_{2},n_{3}}\mid n_{1},n_{2}%
,n_{3}\rangle
\end{equation}
 we end up with 
the eigenvalues
\begin{equation}
E_{n_{1},n_{2},n_{3}}=\hbar\varpi\left(  e^{\varsigma-\rho}n_{1}%
+e^{\kappa-\varsigma}n_{2}+e^{\rho-\kappa}n_{3}+\frac{e^{\varsigma-\rho
}+e^{\kappa-\varsigma}+e^{\rho-\kappa}}{2}\right)
\end{equation}
as well as the normalized 
wavefunctions
\begin{eqnarray}
\psi_{n_{1},n_{2},n_{3}}\left(  q_{1},q_{2},q_{3}\right)   &  =&\left(
\frac{m\varpi}{\pi\hbar}\right)  ^{\frac{3}{4}}\frac{1}{\sqrt{2^{n_{1}%
+n_{2}+n_{3}}n_{1}!n_{2}!n_{3}!}}e^{-\tfrac{m\varpi}{2\hbar}\left(
e^{\varsigma-\rho}q_{1}^{2}+e^{\kappa-\varsigma}q_{2}^{2}+e^{\rho-\kappa}%
q_{3}^{2}\right)  }\nonumber\label{states}\\
&&  \times H_{n_{1}}\left(  \sqrt{\tfrac{m\varpi e^{\varsigma-\rho}}{\hbar}%
}q_{1}\right)  H_{n_{2}}\left(  \sqrt{\tfrac{m\varpi e^{\kappa-\varsigma}%
}{\hbar}}q_{2}\right)  H_{n_{3}}\left(  \sqrt{\tfrac{m\varpi e^{\rho-\kappa}%
}{\hbar}}q_{3}\right)
\end{eqnarray}
where $n_{1},n_{2},n_{3}$ are three integer numbers
and $H_{n_i}$ are Hermite polynomials, with $i=1,2,3$. 
It is clearly seen that \eqref{states} is  tensor product  of three
independent solutions such that each one is corresponding to
a harmonic oscillator in one dimension.

To express \eqref{states}
in terms of old variables $(x_1, x_2, x_3)$ we use the reciprocal transformations
of \eqref{244} to write the new variables $(q_1, q_2, q_3)$ as
\begin{align}
q_{1}  &  =\mu_{1}\cos\theta\cos\phi x_{1}-\mu_{2}\left(  \sin\theta
\sin\varphi+\cos\theta\cos\varphi\sin\phi\right)  x_{2}-\mu_{3}\left(
\sin\theta\cos\varphi-\cos\theta\sin\phi\sin\varphi\right)  x_{3}\label{422}\\
q_{2}  &  =\mu_{1}\sin\phi x_{1}+\mu_{2}\cos\phi\cos\varphi x_{2}-\mu_{3}%
\cos\phi\sin\varphi x_{3}\label{423}\\
q_{3}  &  =\mu_{1}\cos\phi\sin\theta x_{1}+\mu_{2}\left(  \cos\theta
\sin\varphi-\sin\theta\cos\varphi\sin\phi\right)  x_{2}+\mu_{3}\left(
\cos\theta\cos\varphi+\sin\theta\sin\phi\sin\varphi\right)  x_{3} \label{424}%
\end{align}
and then replace in \eqref{states} to
get
the exact  wavefunctions $\psi_{n_{1},n_{2},n_{3}}\left(  x_{1},x_{2},x_{3}\right)$ solutions of the problem of three coupled
harmonic oscillators with quadratic interaction described by the Hamiltonian \eqref{ham1}. {We emphasis that these solutions are general and derived 
without any assumption or approximation.} 



\section{Entanglement in ground state}


To analyze the entanglement of the three-body  system, we first determine
the ground state. Indeed,
from general solutions \eqref{states} together with variable changes (\ref{422}%
-\ref{424}), we  extract the ground state wavefunction
\begin{equation}
\psi_{0,0,0}\left(  x_{1},x_{2},x_{3}\right)  \sim e^{-A\mu_{1}^{2}x_{1}%
^{2}-B\mu_{2}^{2}x_{2}^{2}-C\mu_{3}^{2}x_{3}^{2}+2\mu_{1}\mu_{2}\Gamma
_{12}x_{1}x_{2}+2\mu_{1}\mu_{3}x_{1}x_{3}\Gamma_{13}+2\Gamma_{23}\mu_{2}%
\mu_{3}x_{2}x_{3}}%
\end{equation}
such that all parameters read as
\begin{eqnarray}
A  & = &\alpha\cos^{2}\theta\cos^{2}\phi+\beta\sin^{2}\phi+\gamma\cos^{2}%
\phi\sin^{2}\theta\\
B  & = &\alpha\left(  \sin\theta\sin\varphi+\cos\theta\cos\varphi\sin
\phi\right)  ^{2}+\beta\cos^{2}\phi\cos^{2}\varphi+\gamma\left(  \cos
\theta\sin\varphi-\sin\theta\cos\varphi\sin\phi\right)  ^{2}\\
C  & =& \alpha\left(  \sin\theta\cos\varphi-\cos\theta\sin\phi\sin
\varphi\right)  ^{2}+\beta\cos^{2}\phi\sin^{2}\varphi+\gamma\left(  \cos
\theta\cos\varphi+\sin\theta\sin\phi\sin\varphi\right)  ^{2}\\
\Gamma_{12}  & =&\alpha\cos\theta\cos\phi\left(  \sin\theta\sin\varphi
+\cos\theta\cos\varphi\sin\phi\right)  -\beta\sin\phi\cos\phi\cos
\varphi\nonumber\\
&&  -\gamma\cos\phi\sin\theta\left(  \cos\theta\sin\varphi-\sin\theta
\cos\varphi\sin\phi\right) \\
\Gamma_{13}  & =& -\alpha\cos\theta\cos\phi\left(  -\sin\theta\cos\varphi
+\cos\theta\sin\phi\sin\varphi\right)  +\beta\sin\phi\cos\phi\sin
\varphi\nonumber\\
&&  -\gamma\cos\phi\sin\theta\left(  \cos\theta\cos\varphi+\sin\theta\sin
\phi\sin\varphi\right) \\
\Gamma_{23}  & =& \alpha\left(  \sin\theta\sin\varphi+\cos\theta\cos\varphi
\sin\phi\right)  \left(  -\sin\theta\cos\varphi+\cos\theta\sin\phi\sin
\varphi\right) \\
&&  +\beta\cos\phi\cos\varphi\cos\phi\sin\varphi-\gamma\left(  \cos\theta
\sin\varphi-\sin\theta\cos\varphi\sin\phi\right)  \left(  \cos\theta
\cos\varphi+\sin\theta\sin\phi\sin\varphi\right) \nonumber\\
%
\alpha &=& \tfrac{m\varpi}{2\hbar}e^{\varsigma-\rho},
\qquad \beta=\tfrac{m\varpi}{2\hbar}e^{\kappa-\varsigma},\qquad \gamma=\tfrac{m\varpi
}{2\hbar}e^{\rho-\kappa}. 
\end{eqnarray}
Once the ground state wavefunction corresponding to
our system is obtained, we now return to explicitly determine the reduced
density matrix. Then based on the standard definition%
\begin{equation}
\rho_{\mathsf{red}}^{A}(x_{1},x_{1}^{\prime})=\frac{\int\psi_{0,0,0}\left(
x_{1},x_{2},x_{3}\right)  \psi_{0,0,0}^{\ast}\left(  x_{1}^{\prime}%
,x_{2},x_{3}\right)  dx_{2}dx_{3}}{\int\psi_{0,0,0}\left(  x_{1},x_{2}%
,x_{3}\right)  \psi_{0,0,0}^{\ast}\left(  x_{1},x_{2},x_{3}\right)
dx_{1}dx_{2}dx_{3}} \label{313}%
\end{equation}
we find the result
\begin{equation}
\rho_{\mathsf{red}}^{A}(x_{1},x_{1}^{\prime})=\sqrt{\frac{2L-w}{\pi}%
}e^{-L\left(  x_{1}^{2}+x_{1}^{\prime2}\right)  +wx_{1}x_{1}^{\prime}}%
\end{equation}
and 
the two parameters are given by
\begin{align}
L  &  =\mu_{1}^{2}\left(A-\tfrac{\Gamma_{13}^{2}}{2C}-\tfrac{\left( \Gamma_{12}+\frac{\Gamma_{13}\Gamma_{23}}{C}\right)  ^{2}}{2\left(
B^{2}-\frac{\Gamma_{23}^{2}}{C%
}\right)  }\right)\\
w  &  =\mu_{1}^{2} \left(\tfrac{\Gamma_{13}^{2}}{C}%
+\tfrac{\left(  \Gamma_{12}+\frac{\Gamma_{13}\Gamma_{23}}{C}\right)  ^{2}}{B-\frac
{\Gamma_{23}^{2}}{C}}\right).
\end{align}

Now we are in the final stage to talk about entanglement
of our system. Indeed, the corresponding purity function can be obtained as
\begin{equation}
P=\int\rho_{\mathsf{red}}^{A}(x_{1},x_{1}^{\prime})\rho_{\mathsf{red}}%
^{A}(x_{1}^{\prime},x_{1})dx_{1}dx_{1}^{\prime}=\sqrt{\frac{2L-w}{2L+w}}.
\end{equation}
Replacing different quantities and after straightforward calculation we end up
with the final form of such  function
\begin{eqnarray}
\label{purity}P  &  =& \tfrac{1}{\sqrt{e^{\rho-\varsigma} \cos^{2}\theta\cos^{2}\phi
+e^{\varsigma-\kappa} \sin^{2}\phi +e^{\kappa-\rho}\cos^{2}\phi\sin
^{2}\theta }}\nonumber\\
&&  \times\tfrac{1}{\sqrt{e^{\rho-\varsigma}\left(  \sin\theta\sin\varphi+\cos\theta\cos
\varphi\sin\phi\right)  ^{2}+ e^{\varsigma-\kappa}\cos^{2}\phi\cos^{2}\varphi
+ e^{\kappa-\rho}\left(  \cos\theta\sin\varphi-\sin\theta\cos\varphi
\sin\phi\right)  ^{2}}}\nonumber\\
&&  \times\tfrac{1}{\sqrt{e^{\rho-\varsigma}\left(  -\sin\theta\cos\varphi+\cos\theta\sin\phi
\sin\varphi\right)  ^{2}+e^{\varsigma-\kappa} \cos^{2}\phi\sin^{2}\varphi
+e^{\kappa-\rho} \left(  \cos\theta\cos\varphi+\sin\theta\sin\phi
\sin\varphi\right)  ^{2}}}.
\end{eqnarray}
At this level, we have some comments in order. Indeed
\eqref{purity} is actually depending on a set of six parameters $(\rho
,\varsigma,\kappa,\theta,\varphi,\phi)$ and therefore 
one may consider
different configurations to numerically analyze the behavior of 
our system.
This will be not done here because our concern is to give an exact solution of the problem under consideration. Nevertheless, we can still talk about minimal and maximal values of the purity function to give some ideas about the entanglement of our system.
More precisely,
two situations will be analyzed with respect to the strength of the coupling
parameters $\left(  \varsigma,\rho,\kappa\right)  $, which will allow us to
see how much the present system is entangled. 

We start with the weak coupling
that is characterized by taking the limit $\left(  J_{12},J_{13}%
,J_{23}\right)  \longrightarrow\left(  0,0,0\right)  $ where the angles
$\left(  \theta,\phi,\varphi\right)  \longrightarrow\left(  \theta_{w}%
,\phi_{w},\varphi_{w}\right)  $ and the coupling $\left(  \varsigma
,\rho,\kappa\right)  \longrightarrow\left(  \varsigma_{w},\rho_{w},\kappa
_{w}\right)  $. In this case,  (\ref{j12}-\ref{j23}) and \eqref{177} reduce to the following
\begin{equation}
\left(  \theta_{w},\phi_{w},\varphi_{w}\right)  =\left(  0,0,0\right),\quad
e^{\varsigma_{w}-\rho_{w}}=\tfrac{\omega_{1}}{\left(  \omega_{1}\omega
_{2}\omega_{3}\right)  ^{\frac{1}{3}}},\quad e^{\kappa_{w}-\varsigma_{w}%
}=\tfrac{\omega_{2}}{\left(  \omega_{1}\omega_{2}\omega_{3}\right)  ^{\frac
{1}{3}}},\quad e^{\rho_{w}-\kappa_{w}}=\tfrac{\omega3}{\left(  \omega
_{1}\omega_{2}\omega_{3}\right)  ^{\frac{1}{3}}}%
\end{equation}
which can be implemented into \eqref{purity} to get the maximal value of the purity function
\begin{equation}
P\left(  \varsigma_{w},\rho_{w},\kappa_{w},\theta_{w},\phi_{w},\varphi
_{w}\right)  =1
\end{equation}
 showing that the system is completely separable and therefore there is no
entangled states because of the entropy $S=1-P$.

Now we consider the strong coupling limit 
corresponding to 
the limit $\left(  \varsigma,\rho
,\kappa\right)  \longrightarrow\left(  \varsigma_{s},\rho_{s},\kappa
_{s}\right)  $ and $\left(  \theta,\phi,\varphi\right)  \longrightarrow\left(
\theta_{s},\phi_{s},\varphi_{s}\right)  $. Doing this process to obtain the limit
\begin{equation}
\left(  \varsigma_{s}-\rho_{s},\kappa_{s}-\varsigma_{s},\rho_{s}-\kappa
_{s}\right)  \longrightarrow\left(  \pm\infty,\pm\infty,\pm\infty\right)
\end{equation}
and therefore the purity function \eqref{purity} reduces to the following quantity
\begin{equation}
P\left(  \varsigma_{s},\rho_{s},\kappa_{s},\theta_{s},\phi_{s},\varphi
_{s}\right)  \longrightarrow0
\end{equation}
 telling us 
that our system is maximally entangled because of $S=1$. This summarizes that there
are two extremely values of the purity function those could be reached as long as the
coupling parameters take small or large values.


\section{Limiting cases}


Now we will see how to derive some results already know in literature, which
concern three limiting cases to distinguish in terms of the coupling
parameters where the first one will be treated below and tow remaining will be summarized in Appendix {\bf B}. We emphasis that  all such cases will give the same results 
but the main differences are how to fix the physical parameters
and choose coordinate variables.
To get the solutions of two coupled harmonic oscillators in variables $\left(
x_{1},x_{2}\right)  $
we simply require the limits $D_{13},D_{23}\longrightarrow0$, which correspond
to
$J_{13},J_{23}\longrightarrow0$. These operations restrict the Hamiltonian
$H_{1}$ to the following
\begin{equation}
H_{1}\longrightarrow H_{0}+\frac{p_{3}^{2}}{2m_{3}}+\frac{1}{2}m_{3}\omega
_{3}^{2}x_{3}^{2}%
\end{equation}
where $H_{0}$ is
the Hamiltonian of the two coupled harmonic oscillators in $\left(
x_{1},x_{2}\right)  $ variables
\begin{equation}
H_{0}=\frac{p_{1}^{2}}{2m_{1}}+\frac{p_{2}^{2}}{2m_{2}}+\frac{1}{2}m_{1}%
\omega_{1}^{2}x_{1}^{2}+\frac{1}{2}m_{2}\omega_{2}^{2}x_{2}^{2}+\frac{1}%
{2}D_{12}x_{1}x_{2}.
\end{equation}
By taking $J_{13},J_{23}\longrightarrow0$ in (\ref{j13}-\ref{j23}) we obtain
$\theta\longrightarrow0, \varphi\longrightarrow0$ and
\begin{align}
\omega_{1}^{2}  &  \longrightarrow\Sigma_{1}^{2}\cos^{2}\phi+\Sigma_{2}%
^{2}\sin^{2}\phi=\frac{\Sigma_{1}^{2}+\Sigma_{2}^{2}}{2}+\frac{\Sigma_{1}%
^{2}-\Sigma_{2}^{2}}{2}\cos2\phi\\
\quad\omega_{2}^{2}  &  \longrightarrow\Sigma_{2}^{2}\cos^{2}\phi+\Sigma
_{1}^{2}\sin^{2}\phi=\frac{\Sigma_{1}^{2}+\Sigma_{2}^{2}}{2}-\frac{\Sigma
_{1}^{2}-\Sigma_{2}^{2}}{2}\cos2\phi,\\
\omega_{3}^{2}  &  \longrightarrow\Sigma_{3}^{2}\\
\quad J_{12}  &  \longrightarrow-\tfrac{\Sigma_{1}^{2}-\Sigma_{2}^{2}}{2}%
\sin2\phi
\end{align}
showing that the reciprocal expressions take the forms
\begin{align}
\Sigma_{1}^{2}  &  =\tfrac{\omega_{1}^{2}+\omega_{2}^{2}+\frac{\omega_{1}%
^{2}-\omega_{2}^{2}}{\cos2\phi}}{2}=\tfrac{\omega_{1}^{2}+\omega_{2}^{2}%
+\sqrt{\left(  \omega_{1}^{2}-\omega_{2}^{2}\right)  ^{2}+J_{12}^{2}}}%
{2}=k_{12}e^{e^{+2\eta_{12}}}\\
\Sigma_{1}^{2}  &  =\tfrac{\omega_{1}^{2}+\omega_{2}^{2}-\frac{\omega_{1}%
^{2}-\omega_{2}^{2}}{\cos2\phi}}{2}=\tfrac{\omega_{1}^{2}+\omega_{2}^{2}%
-\sqrt{\left(  \omega_{1}^{2}-\omega_{2}^{2}\right)  ^{2}+J_{12}^{2}}}%
{2}=k_{12}e^{-2\eta_{12}}\\
\Sigma_{3}^{2}  &  =\omega_{3}^{2},\varpi=\left(  \Sigma_{1}\Sigma_{2}%
\Sigma_{3}\right)  ^{\frac{1}{3}}\longrightarrow\left(  k_{12}\omega_{3}\right)
^{\frac{1}{3}}\\
e^{\varsigma-\rho}  &  =\frac{\sqrt{k_{12}}e^{e^{+\eta_{12}}}}{\left(  k_{12}\omega
_{3}\right)  ^{\frac{1}{3}}},\qquad e^{\kappa-\varsigma}=\frac{\sqrt
{k_{12}}e^{e^{-\eta_{12}}}}{\left(  k_{12}\omega_{3}\right)  ^{\frac{1}{3}}},\qquad
e^{\rho-\kappa}=\tfrac{\omega_{3}}{\left(  k_{12}\omega_{3}\right)  ^{\frac{1}{3}}}%
\end{align}
where we have set
\begin{equation}
e^{\pm2\eta_{12}}=\tfrac{\omega_{1}^{2}+\omega_{2}^{2}\pm\sqrt{\left(  \omega
_{1}^{2}-\omega_{2}^{2}\right)  ^{2}+J_{12}^{2}}}{2k_{12}},\qquad k_{12}=\sqrt
{\omega_{1}^{2}\omega_{2}^{2}-\tfrac{J_{12}^{2}}{4}}.
\end{equation}
It is clearly seen that the above sets are those used in our previous work
\cite{Jellal11} to decouple the problem of two harmonic oscillators in
$\left(  x_{1},x_{2}\right)  $ variables
\begin{equation}
H=\left(  \frac{p_{1}^{2}}{2m}+\frac{p_{2}^{2}}{2m}+\frac{m}{2}k_{12}e^{2\eta_{12}}%
q_{1}^{2}+\frac{m}{2}k_{12}e^{-2\eta_{12}}q_{2}^{2}\right)  +\frac{p_{3}^{2}}{2m_{3}%
}+\frac{1}{2}m\omega_{3}^{2}q_{3}^{2}%
\end{equation}
whose eigenvalues and the eigenstates are given by
\begin{equation}
E_{n_{1},n_{2},n_{3}}=\hbar\sqrt{k_{12}}\left(  e^{\eta_{12}}n_{1}+e^{-\eta_{12}}n_{2}%
+\cosh\eta_{12}\right)  +\hbar\omega_{3}\left(  n_{3}+\frac{1}{2}\right)
\end{equation}%
\begin{eqnarray}
\psi_{n_{1},n_{2},n_{3}}\left(  x_{1},x_{2},x_{3}\right)   &  =&\frac{\left(
\frac{m_{3}\omega_{3}}{\pi\hbar}\right)  ^{\frac{1}{4}}\left(  \frac{m\sqrt
{k_{12}}}{\pi\hbar}\right)  ^{\frac{1}{2}}}{\sqrt{2^{n_{1}+n_{2}+n_{3}}n_{1}%
!n_{2}!n_{3}!}}e^{-\tfrac{m}{2\hbar}\left(  \sqrt{k_{12}}e^{\eta_{12}}q_{1}^{2}+\sqrt
{k_{12}}e^{-\eta_{12}}q_{2}^{2}+\omega_{3}q_{3}^{2}\right)  }\nonumber\\
&&  \times H_{n_{1}}\left(  \sqrt{\tfrac{m\sqrt{k_{12}}e^{\eta_{12}}}{\hbar}}%
q_{1}\right)  H_{n_{2}}\left(  \sqrt{\tfrac{m\sqrt{k_{12}}e^{-\eta_{12}}}{\hbar}}%
q_{2}\right)  H_{n_{3}}\left(  \sqrt{\tfrac{m\omega_{3}}{\hbar}}q_{3}\right)
\end{eqnarray}
with the variables
\begin{equation}
q_{1}=\mu_{1}\cos\phi x_{1}-\mu_{2}\sin\phi x_{2},\qquad q_{2}=\mu_{1}\sin\phi
x_{1}+\mu_{2}\cos\phi x_{2},\qquad q_{3}=\mu_{3}x_{3}.
\end{equation}
The corresponding purity function can be derived from \eqref{purity} as limiting case%
\begin{align}
P\left(  \rho,\varsigma,\kappa,\theta=0,\varphi=0,\phi\right)   &  =\tfrac
{1}{\sqrt{\left(  e^{\rho-\varsigma}\cos^{2}\phi +e^{\varsigma
-\kappa}\sin^{2}\phi \right)  \left( e^{\rho-\varsigma} \sin^{2}\phi +e^{\varsigma-\kappa} \cos^{2}\phi
\right)  e^{\kappa-\rho}}}\nonumber\\
&  =\tfrac{1}{\sqrt{\left( e^{-\eta_{12}} \cos^{2}\phi 
+e^{+\eta_{12}} \sin^{2}\phi \right)  \left( e^{-\eta_{12}} \sin^{2}\phi +
e^{+\eta_{12}} \cos^{2}\phi \right)  }%
}=P_{0,0}\left(  \eta_{12},\phi\right)
\end{align}
which coincides exactly with that obtained in our previous work
\cite{Jellal11}.  The two other limiting cases are discussed in 
{\bf Appendix B} and the obtained results are similar except that
the configurations of physical parameters together with variable coordinates are not the same.


\section{Conclusion}


We have studied the problem of three coupled harmonic oscillators involving general
coupling between coordinates. 
In doing so, 
different transformations have been introduced to finally
end up with 
the solutions of the energy spectrum. More precisely, the representation theory of the group $SU(3)$ was employed to get a diagonalizable Hamiltonian
describing three decoupled harmonic oscillators. Later on, the reciprocal transformations
were used to express the general solutions of the interacting system in terms of the initial coordinates. The obtained results are general and derived 
without making use of any assumption or approximation.

Subsequently, focused on the ground state wavefunction we have calculated
the corresponding reduced density matrix. This was used to explicitly determine
the purity function in terms of different physical parameters 
of three coupled harmonic oscillators as well as obtain its minimal and maximal values. To check the validity of results, we have
inspected three limiting cases, which have been done by realizing one among three oscillators.  In each case we have established the corresponding conditions as well as the convenient variable changes. In all cases, we have obtained the same entanglement as
has been reported in 
\cite{Jellal11}.

{The present work will not remain at this stage because  we plan to investigate other issues using the obtained results. Indeed, first question will deal with the corresponding thermodynamic properties 
and second one will concern the dynamics of the entanglement by considering frequencies time dependent. All these questions and related matters are actually under consideration.
}

\section*{Acknowledgments}

AM acknowledges the Deanship of Scientific Research at King Faisal
University for the financial support under Nasher Track (Grant No. 186262).
The generous support provided by the Saudi Center for Theoretical
Physics (SCTP) is highly appreciated by AJ.

\section*{\textbf{Appendix A: $SU(3)$ algebra}}


We recall some mathematical tools related to {Lie group $SU(3)$}, which have been used in our work. Indeed,
the generators of $SU(3)$ are given by \cite{Gelman matrices}
\begin{eqnarray}
&& \lambda_{1}    =\left(
\begin{matrix}
0 & 1 & 0\\
1 & 0 & 0\\
0 & 0 & 0
\end{matrix}
\right)  ,\quad\lambda_{2}=\left(
\begin{matrix}
0 & -i & 0\\
i & 0 & 0\\
0 & 0 & 0
\end{matrix}
\right)  ,\qquad\lambda_{3}=\left(
\begin{matrix}
1 & 0 & 0\\
0 & -1 & 0\\
0 & 0 & 0
\end{matrix}
\right)\nonumber \\  
&& \lambda_{4}=\left(
\begin{matrix}
0 & 0 & 1\\
0 & 0 & 0\\
1 & 0 & 0
\end{matrix}
\right), \qquad 
\lambda_{5}    =\left(
\begin{matrix}
0 & 0 & -i\\
0 & 0 & 0\\
i & 0 & 0
\end{matrix}
\right)  ,\quad\lambda_{6}=\left(
\begin{matrix}
0 & 0 & 0\\
0 & 0 & 1\\
0 & 1 & 0
\end{matrix}
\right) \\
&& \lambda_{7}=\left(
\begin{matrix}
0 & 0 & 0\\
0 & 0 & -i\\
0 & i & 0
\end{matrix}
\right)  ,\quad\lambda_{8}=\frac{1}{\sqrt{3}}
\left(
\begin{matrix}
1 & 0 & 0\\
0 & 1 & 0\\
0 & 0 & -2
\end{matrix}
\right)
\nonumber
\end{eqnarray}
where  the Gell-Mann matrices $\lambda_{i}$, that are 
analog of the Pauli matrices for the group $SU(2)$, satisfy the $SU(3)$ commutation
relations
\begin{equation}
\left[  \lambda_{j},\lambda_{k}\right]  =2i%
{\textstyle\sum_{l}}
f^{jkl}\lambda_{l} 
\end{equation}
and  the structure constants $f^{ijk}$ of the Lie algebra are given by
\begin{eqnarray}
&& f^{123}=1 \nonumber \\
&& f^{147}=-f^{156}=f^{246}=f^{257}=f^{345}=-f^{367}=\frac{1}%
{2}\\
&& f^{458}=f^{678}=\frac{\sqrt{3}}{2}.\nonumber
\end{eqnarray}
We can check the useful identities%
\begin{eqnarray}
&& e^{-i\phi\lambda_{2}}\lambda_{6}e^{+i\phi\lambda_{2}}=\cos\phi\lambda
_{6}-\sin\phi\lambda_{4}
\\
&&
e^{-i\phi\lambda_{2}}\lambda
_{4}e^{+i\phi\lambda_{2}}=\cos\phi\lambda_{4}+\sin\phi\lambda_{6}
\\
&&
e^{-i\phi\lambda_{2}}\lambda_{1}e^{+i\phi\lambda_{2}}=\text{diag}\left(
-\sin2\phi,\sin2\phi,0\right)  +\cos\left(  2\phi\right)  \lambda_{1} \\
&&
e^{-i\phi\lambda_{2}}\text{diag}\left(  a,b,c\right)  e^{+i\phi\lambda_{2}%
}=\text{diag}\left(  a\cos^{2}\phi+b\sin^{2}\phi,b\cos^{2}\phi+a\sin^{2}%
\phi,c\right)  +\frac{a-b}{2}\sin2\phi\lambda_{1} \\
&&
e^{-i\theta\lambda_{5}}\lambda_{6}e^{i\theta\lambda_{5}}=\cos\theta
\lambda_{6}-\sin\theta\lambda_{1}\\
&& e^{-i\theta\lambda_{5}}\lambda
_{1}e^{i\theta\lambda_{5}}=\cos\theta\lambda_{1}+\sin\theta\lambda_{6} \\
&& e^{-i\theta\lambda_{5}}\lambda_{4}e^{i\theta\lambda_{5}}=\text{diag}\left(
-\sin2\theta,0,\sin2\theta\right)  +\cos2\theta\lambda_{4} \\
&& e^{-i\theta\lambda_{5}}\text{diag}\left(  a,b,c\right)  e^{i\theta\lambda
_{5}}=\text{diag}\left(  a\cos^{2}\theta+c\sin^{2}\theta,b,c\cos^{2}%
\theta+a\sin^{2}\theta\right)  \allowbreak+\frac{a-c}{2}\sin2\theta\lambda
_{4} \\
&& e^{-i\varphi\lambda_{7}}\lambda_{1}e^{+i\varphi\lambda_{7}}=\cos
\varphi\lambda_{1}+\sin\varphi\lambda_{4}\\
&& e^{-i\varphi\lambda_{7}%
}\lambda_{4}e^{+i\varphi\lambda_{7}}=\cos\varphi\lambda_{4}-\sin\varphi
\lambda_{1}\\
&& e^{-i\varphi\lambda_{7}}\lambda_{6}e^{+i\varphi\lambda_{7}}=\text{diag}%
\left(  0,-\sin2\varphi,\sin2\varphi\right)  +\cos2\varphi\lambda_{6} \\
&& e^{-i\varphi\lambda_{7}}\text{diag}\left(  a,b,c\right)  e^{+i\varphi
\lambda_{7}}=\text{diag}\left(  a,b\cos^{2}\varphi+c\sin^{2}\varphi,c\cos
^{2}\varphi+b\sin^{2}\varphi\right)  \allowbreak+\frac{b-c}{2}\sin
2\varphi\lambda_{6}.
\end{eqnarray}

\section*{\textbf{Appendix B: More couplings}}

{For the two coupled harmonic oscillators in variables $\left(  x_{1}%
,x_{3}\right)  $, we have the limits $D_{12},D_{23}\longrightarrow0$ implying
that $J_{12},J_{23}\longrightarrow0$, then from \eqref{j12} and \eqref{j23}
one gets $\varphi\longrightarrow0,\phi\longrightarrow0$. Using 
(\ref{o11}-\ref{o33}) and \eqref{j13}, we obtain}
\begin{align}
\omega_{1}^{2}  &  \longrightarrow\Sigma_{1}^{2}\cos^{2}\theta+\Sigma_{3}%
^{2}\sin^{2}\theta=\tfrac{\Sigma_{1}^{2}+\Sigma_{3}^{2}}{2}+\tfrac{\Sigma
_{1}^{2}-\Sigma_{3}^{2}}{2}\cos2\theta\\
\quad\omega_{2}^{2}  &  \longrightarrow\Sigma_{2}^{2}\\
\omega_{3}^{2}  &  \longrightarrow\Sigma_{3}^{2}\cos^{2}\theta+\Sigma_{1}%
^{2}\sin^{2}\theta=\tfrac{\Sigma_{1}^{2}+\Sigma_{3}^{2}}{2}-\tfrac{\Sigma
_{1}^{2}-\Sigma_{3}^{2}}{2}\cos2\theta\\
\quad J_{13}  &  \longrightarrow-\tfrac{\Sigma_{1}^{2}-\Sigma_{3}^{2}}{2}%
\sin2\theta
\end{align}
and the variables take the form
\begin{equation}
q_{1}=\mu_{1}\cos\theta x_{1}-\mu_{3}\sin\theta x_{3},\qquad q_{2}=\mu
_{2}x_{2},\qquad q_{3}=\mu_{1}\sin\theta x_{1}+\mu_{3}\cos\theta x_{3}
\end{equation}
{Replacing in \eqref{purity} to
derive 
purity function} 
\begin{align}
P\left(  \rho,\varsigma,\kappa,\theta,\varphi=0,\phi=0\right)   &  =\tfrac
{1}{\sqrt{\left( e^{\rho-\varsigma} \cos^{2}\theta +e^{\kappa-\rho}\sin^{2}\theta
\right)  e^{\varsigma-\kappa}\left( e^{\rho-\varsigma} \sin^{2}\theta
+e^{\kappa-\rho}\cos^{2}\theta \right)  }}\nonumber\\
&  =\tfrac{1}{\sqrt{\left( e^{-\eta_{13}} \cos^{2}\theta +e^{+\eta_{13}
}\sin^{2}\theta \right)  \left( e^{-\eta_{13}} \sin^{2}\theta +e^{+\eta_{13}}\cos^{2}\theta \right)  }%
}=P_{0,0}\left(  \eta_{13},\theta\right).
\end{align}

{For the two coupled harmonic oscillator in variables $\left(
x_{2},x_{3}\right)  $, we put $D_{12},D_{13}\longrightarrow0$ giving
$J_{12},J_{13}\longrightarrow0$, then
 (\ref{j12}-\ref{j13}) allow to obtain $\theta\longrightarrow
0,\phi\longrightarrow0$. From (\ref{o11}-\ref{o33}) and \eqref{j23}, we
get}
\begin{align}
\omega_{1}^{2}  &  \longrightarrow\Sigma_{1}^{2}\\
\quad\omega_{2}^{2}  &  \longrightarrow\Sigma_{2}^{2}\cos^{2}\varphi
+\Sigma_{3}^{2}\sin^{2}\varphi=\tfrac{\Sigma_{2}^{2}+\Sigma_{3}^{2}}{2}%
+\tfrac{\Sigma_{2}^{2}-\Sigma_{3}^{2}}{2}\cos2\varphi\\
\omega_{3}^{2}  &  \longrightarrow\Sigma_{3}^{2}\cos^{2}\varphi+\Sigma_{2}%
^{2}\sin^{2}\varphi=\tfrac{\Sigma_{2}^{2}+\Sigma_{3}^{2}}{2}-\tfrac{\Sigma
_{2}^{2}-\Sigma_{3}^{2}}{2}\cos2\varphi\\
\quad J_{23}  &  \longrightarrow-\tfrac{\Sigma_{2}^{2}-\Sigma_{3}^{2}}{2}%
\sin2\varphi.
\end{align}
The corresponding variables are given by
\begin{equation}
q_{1}=\mu_{1}x_{1},\qquad q_{2}=\mu_{2}\cos\varphi x_{2}-\mu_{3}\sin\varphi
x_{3},\qquad q_{3}=\mu_{2}\sin\varphi x_{2}+\mu_{3}\cos\varphi x_{3} 
\end{equation}
as well as the
purity
\begin{align}
P\left(  \rho,\varsigma,\kappa,\theta=0,\varphi,\phi=0\right)   &  =\tfrac
{1}{\sqrt{e^{\rho-\varsigma} \left( e^{\varsigma-\kappa}  \cos^{2}\varphi %
+ e^{\kappa-\rho}\sin^{2}\varphi \right)  \left( e^{\varsigma-\kappa} \sin^{2}\varphi
+e^{\kappa-\rho} \cos^{2}\varphi \right)  }}\nonumber\\
&  =\tfrac{1}{\sqrt{\left(  e^{-\eta_{23}}\cos^{2}\varphi +e^{+\eta_{23}}\sin^{2}\varphi
\right)  \left( e^{-\eta_{23}} \sin^{2}\varphi +e^{+\eta_{23}} \cos^{2}\varphi 
\right)  }}=P_{0,0}\left(  \eta_{23},\varphi\right).
\end{align}

\end{document}